\documentclass[12pt,a4paper]{article}

\usepackage[english]{babel}
\usepackage{fancyhdr}
\usepackage{amsmath}
\usepackage{amssymb}
\usepackage{amsfonts}
\usepackage{psfrag}
\usepackage{inputenc}
\usepackage{graphicx,wrapfig}
\usepackage[bf,footnotesize]{caption2}
\usepackage{cite}

\usepackage{bm}
\usepackage[dvipsnames]{xcolor}
\usepackage[colorlinks=True, citecolor=blue, linkcolor=blue, urlcolor=blue,linktocpage]{hyperref}
\usepackage[sort&compress,numbers, merge]{natbib}

\newcommand{\beq}{\begin{equation}}
\newcommand{\eeq}{\end{equation}}
\newcommand{\bea}{\begin{eqnarray}}
\newcommand{\eea}{\end{eqnarray}}

\newcommand{\mv}[1]{\langle #1 \rangle}

\begin{document}

\begin{titlepage}
	\begin{center}
	
		\vspace{2.0cm}
		{\Large\bf 
			Constraining a stochastic variation of the gravitational coupling with binary systems}
			
		\vspace{1.0cm}
		\renewcommand{\thefootnote}{\fnsymbol{footnote}}
		{\small \bf 
			Francisco D. Mazzitelli$^{a}$\, \footnote{E-mail:
				\href{mailto:fdmazzi@cab.cnea.gov.ar}{fdmazzi@cab.cnea.gov.ar}} and
			Leonardo G. Trombetta$^{b}$\, \footnote{E-mail:
				\href{mailto:trombetta@fzu.cz}{trombetta@fzu.cz}}
		}
		
		\vspace{0.7cm}
		{\it\footnotesize
			${}^a$Centro At\'omico Bariloche and Instituto Balseiro, Comisi\'on Nacional de Energ\'\i a At\'omica, 8400 Bariloche, Argentina\\
			${}^b$CEICO, Institute of Physics of the Czech Academy of Sciences, Na Slovance 1999/2, 182 21, Prague 8, Czechia
		}
		
		\vspace{0.9cm}
		\abstract{\noindent We consider the effect of stochastic fluctuations of the gravitational coupling $G$ on the evolution of binary systems. We work at an elementary level, in the Newtonian limit, and focus mainly on laser ranging.  We show that, due to cumulative effects, observational data may be used to put bounds on the stochastic fluctuations. We also reanalyze previous results on the implications of stochastic fluctuations of $G$ on cosmological models.   
		}
	\end{center}
\end{titlepage}
\renewcommand{\thefootnote}{\arabic{footnote}}
\setcounter{footnote}{0}

 
\section{Introduction}

Revealing the quantum nature of gravity is a difficult task that nowadays may be within experimental capabilities \cite{Carney:2018ofe}. General Relativity can be thought of as the low energy limit of a quantum field theory  in which the background spacetime metric is considered classical,  and the fluctuations around it are quantized perturbatively \cite{Donoghue:1994dn,Donoghue:2022eay}. This semiclassical theory of gravity is the quantum field theory of gravitons in a curved spacetime \cite{Birrell:1982ix,Parker:2009uva}, and the nonrenormalizability is harmless at low energies. The challenge is to find the appropriate experiments or astrophysical observations that could demonstrate the existence of gravitons.

Effective field theories have their paradigm in quantum Brownian motion \cite{Hu:1991di}. After integrating out the environmental degrees of freedom, a heavy Brownian particle will satisfy a Langevin equation, that is, the effective dynamics will contain both dissipation and noise. While for gravity in most cases the analysis has been circumscribed to dissipative aspects, it is well known that both, dissipation and noise, come together, and the resulting effective field theory is known as stochastic semiclassical gravity \cite{Hu:2020luk}.

Recently, it has been proposed that quantum aspects of gravity could be revealed by interferometric observations: the quantum effects on the geodesics are not only of dissipative character, but should also include a noise induced by the gravitons \cite{Parikh:2020nrd,Parikh:2020fhy}. The magnitude of the effect depends strongly on the quantum state assumed for the gravitons. This proposal has triggered several works devoted to the analysis of the impact of quantum noise on geodesic deviation and tidal forces, as well as on the impact of the different quantum states of the gravitational waves on the interferometric signals \cite{Parikh:2020nrd,Parikh:2020fhy,Haba:2020jqs,Cho:2021gvg,Cho:2023dmh,Chawla:2021lop}. The brand new idea is to go beyond the previous works where the  effects of gravitons have been considered to estimate corrections to the Newtonian potential and the geodesics of test masses \cite{Dalvit:1997yc,Dalvit:1999wd,dePaulaNetto:2021axj}, now including  the unavoidable stochastic consequences of the fluctuations of the metric. Similar ideas have been pursued before  in cosmological scenarios \cite{Wang:2017oiy,Lozano:2020xga}. 

Technically, the integration of the quantum fluctuations of the metric produces a nonlocal Feynman-Vernon influence functional for the classical variables. Dissipative/noise effects are encoded in the real/imaginary part of the influence action, and are connected by the fluctuation-dissipation relation \cite{Calzetta:2008iqa}. In the absence of a full quantum theory of gravity, several nonlocal effective actions have been proposed ``phenomenologically" to analyze the influence of quantum fluctuations of the metric on astrophysical and cosmological effects \cite{Belgacem:2017cqo}. Many  works in this context are intended to determine eventual observational consequences which, in the quantum Brownian motion language, are produced by the dissipative effects. The analysis of the effects of the noise are more rare.

In the same vein, it has been speculated that
integrating out quantum gravitational degrees of freedom could induce stochasticity in the macroscopic gravitational coupling constant $G$, that is, a noise that enters multiplicatively in the effective Einstein equations. In Ref.~\cite{deCesare:2016dnp} a purely time-dependent noisy contribution to $G$ has been considered as a toy model for such effects and its impact in the cosmological evolution has been studied phenomenologically. It was pointed out there that a cosmological evolution consistent with the observed current accelerated expansion of the Universe is possible in this context without the need for a cosmological constant to be introduced in the action by hand. 

Such a possibility is very interesting. However as the observable Universe represents only a single realization of this stochastic process, it is difficult to assess its likelihood. It is then important to study the effects of such stochastic variations of the gravitational coupling in other smaller-scale systems as a mean to assess the viability of such a proposal by opening the possibility of a statistical study. Beyond this cosmological motivation, it can be interesting on its own to characterize the effects of a stochastically varying gravitational coupling on other scenarios.

Binary systems are plentiful and their orbital parameters have in many cases been measured with extreme precision over long periods of time. This includes pulsar timing measurements as well as solar-system planetary ephemerides and the Lunar Laser Ranging (LLR) experiment  \cite{Murphy_2013}. These systems have long been used for the study of gravitational physics to constrain possible deviations from General Relativity  \cite{Will:2014kxa}, as well as looking for expected effects produced by  the stochastic gravitational wave background \cite{Blas:2021mpc}. Characterizing the effect of other sources of stochasticity is important for these efforts. 

At this point it is interesting to remark that, although our main motivation comes from quantum gravity and the theory of quantum open systems, a closely related classical situation is that of the Kepler problem with time-dependent mass, the so-called Gylden's problem \cite{Deprit, Abad_2020}, proposed originally to describe the secular acceleration of the Moon’s longitude. In general, this problem has been investigated under deterministic time dependence. Our results will apply to the case of stochastic variations of the mass.

 In this paper we will consider as a toy model a purely time-dependent $G$ that varies stochastically. Following Ref.~\cite{deCesare:2016dnp}, the gravitational coupling $G$ is assumed to be subject to stochastic fluctuations of the form
\begin{equation} \label{stochastic-G}
 G(t) = \bar{G} (1 + \sigma \xi(t)),
\end{equation}
where $\sigma$ has dimensions of $[T]^{1/2}$ and gives a measure of the intensity of these fluctuations, while $\xi(t)$ stands for the noise and in our convention has dimensions $[T]^{-1/2}$. The statistical properties of the noise are given by its mean value,
\begin{equation} \label{mean-noise}
 \langle \xi(t) \rangle = 0,
\end{equation}
and its self-correlation function at different times,
\begin{equation} \label{correlation}
 \langle \xi(t) \, \xi(t') \rangle = R(t-t').
\end{equation}
Here $R(t)$ is an even function, it has dimensions of $[T]^{-1}$ and should be understood as a distribution. A simple example is the white-noise correlation $R(t-t') = \delta(t-t')$, as assumed in Ref.~\cite{deCesare:2016dnp}, though here we will keep it general for the time being. Equations \eqref{mean-noise} and \eqref{correlation} imply that the gravitational coupling
\begin{eqnarray}
    \langle G(t) \rangle &=& \bar{G}, \\
    \text{Var}(G(t)) &=& \sigma^2 \bar{G}^2 R(0),
\end{eqnarray}
meaning that in this model the value of the gravitational coupling, though fluctuating, remains bounded over time as long as $R(0)$ is finite.  This kind of behavior will not be picked up by experiments looking for a cumulative effect like $G(t) = G_0 + \dot{G} (t - t_0)$. Instead, the stochastic fluctuations of $G(t)$ will need to be studied statistically over a large number of observations of a single system, and/or over a large number of systems that are sensitive to these effects. 

The paper will be organized as follows. In Sec.~\ref{sec:cosmo} we consider the cosmological scenario of Ref.~\cite{deCesare:2016dnp}, and study the stochastic effects on the Hubble parameter by solving perturbatively the Friedmann equations and computing its mean value and variance. We support our results with numerical simulations, and emphasize the strong dependence of the results with the initial conditions. 
In Sec.~\ref{sec:binary} we study the effects on binary systems. We work in a Newtonian approach and, for simplicity, consider perturbations around a circular orbit, which at linear level behave as a harmonic oscillator with both stochastic frequency and driver. We then compute, in Sec~\ref{sec:MSA}, the stochastic mean value and variance of the perturbations using a stochastic generalization of the multiple scale analysis (MSA) \cite{bender78:AMM}. Later in Sec.~\ref{sec:constraints} we use numerical simulations to assess the role of nonlinearities and to set bounds on the characteristics of the noise.  Finally,  in Sec.~\ref{sec:disc} we discuss our results, paying particular attention to the eventual influence that the bounds on the noise coming from LLR may have on the cosmological scenarios with stochastic $G$. We also comment about future prospects. The Appendix contains some details of the MSA calculations.

\section{Stochastic effects on the cosmological evolution}\label{sec:cosmo}

In order to study the effects of a stochastically varying gravitational coupling $G$ on the cosmological evolution, we first need to look into the impact of a time-dependent one in the Friedmann equations for a homogeneous and isotropic universe. Here we are assuming variations only in time, not in space. Then we can incorporate the effect of noise by promoting $G$ to a stochastic variable with a given mean value and correlations, leading to Langevin-like equations. In this first part we will follow closely Ref.~\cite{deCesare:2016dnp}.

We begin by considering the standard Friedmann equations with matter (with energy density $\rho$ and vanishing pressure $p = 0$) plus a time-dependent cosmological ``constant" $\Lambda(t)$,
\begin{eqnarray}
    H^2 &=& \frac{8\pi G}{3} \, \rho + \frac{\Lambda}{3} = \frac{8\pi G}{3} \, (\rho + \lambda), \label{F1} \\
    \dot{H} + H^2 &=& - \frac{4\pi}{3} G \, \rho + \frac{\Lambda}{3} = - \frac{4\pi}{3} G \, (\rho -2\lambda), \label{F2}
\end{eqnarray}
where $H$ is the Hubble parameter, and in the second set of equalities we have defined the vacuum energy $\lambda(t)$ through $\Lambda(t) = 8\pi G(t) \lambda(t)$. It is useful to subtract Eq. \eqref{F1} from Eq. \eqref{F2}, obtaining
\begin{equation}
    \dot{H} = - 4\pi G \, \rho, \label{F3}
\end{equation}
which removes the explicit dependence on $\lambda(t)$. Letting the gravitational coupling be time-dependent $G(t)$, the usual covariant conservation law takes the following form
\begin{equation} \label{cons}
    \dot{\rho} + \left(3H + \frac{\dot{G}}{G} \right) \, \rho + \frac{\dot{\Lambda}}{8\pi G} = 0.    
\end{equation}
This equation describes the eventual transfer of energy between the matter and gravitational sectors. As discussed in Ref.~\cite{Fritzsch:2012qc}, there are many possibilities, the simplest one being to assume the standard conservation for matter and a dynamical interplay between the time dependence of $G$ and $\Lambda$. Other choices would require us to further specify a model for the coupling between ordinary matter and the new degrees of freedom responsible for this time dependence. 

Here we will make the simplifying assumption that regular matter is independently conserved and satisfies the standard conservation equation:
\begin{equation} \label{cons-rho}
    \dot{\rho} + 3H \, \rho = 0,
\end{equation}
which, when combined with Eq. \eqref{cons} gives 
\begin{equation}
    (\rho + \lambda) \, \dot{G} + G \, \dot{\lambda} = 0.
\end{equation}
This implies that a time-dependent gravitational coupling $G(t)$ must be balanced by a time dependent vacuum energy $\lambda(t)$ (or vice versa). The above equation is better expressed as
\begin{equation}
    \rho \, \dot{G} + \frac{d}{dt} \left( G \, \lambda \right) = 0,
\end{equation}
which we can integrate to
\begin{eqnarray}
    G(t) \, \lambda(t)  &=& G_i \, \lambda_i - \int_{t_i}^t dt' \, \rho(t') \, \dot{G}(t') \notag \\ 
                        &=& G_i \, (\rho_i + \lambda_i) - \rho(t) G(t) - \int_{t_i}^t dt' \, G(t') \, \dot{\rho}(t'),
\end{eqnarray}
with the initial conditions denoted with the subindex $i$. Inserting this expression back into Eq. \eqref{F1} allows us to obtain an expression for $H^2$ in terms of only $G(t)$ and $\rho(t)$,
\begin{equation}
    H^2 = \frac{8\pi}{3} \left[ G_i \, (\rho_i + \lambda_i) - \int_{t_i}^t dt' \,  G(t') \, \dot{\rho}(t') \right].
\end{equation}
Finally, notice that from this last equation and Eq. \eqref{cons-rho} it follows that the initial conditions are not all independent, but rather they must satisfy
\begin{equation} \label{IC-consistency}
    H_i^2 = \frac{8\pi}{3} G_i \, (\rho_i + \lambda_i) \simeq \frac{8\pi}{3} G_i \, \rho_i, 
\end{equation}
where we are assuming in the last equality that $\lambda_i \ll \rho_i$, such that initially there is no vacuum energy.

We can now proceed to incorporate the stochastic effects by considering that $G(t)$ varies stochastically according to Eq.~\eqref{stochastic-G}, with the statistical properties of the noise given by Eqs.~\eqref{mean-noise} and \eqref{correlation}. Replacing these into \eqref{F3} and \eqref{cons-rho}, we obtain the corresponding Friedmann-Langevin equations,
\begin{subequations} \label{langevin-cosmo}
\begin{align}
\dot{\rho} &= - 3H \rho, \label{langevin-rho} \\
\dot{H} &= - 4\pi \bar{G} \rho (1 + \sigma \xi) . \label{langevin-H} 
\end{align}
\end{subequations}
The first of these is actually the same as in the deterministic case; however, the second equation ensures that $H$ is now a stochastic variable, and therefore this also permeates to $\rho$ via the first one.

Let us write $H=H_D+\delta H$ and $\rho=\rho_D+\delta\rho$ where the subindex $D$ denotes the deterministic solution of the above equations.  Consider a specific realization of the stochastic process where $H > H_D$ ($\delta H > 0$), Eq. \eqref{langevin-rho} tells us that $\rho \to 0$ more quickly than $\rho_D$, and therefore, according to Eq. \eqref{langevin-H}, $\dot{H} \to 0$ faster than $\dot{H}_D$, so it is expected in this case that $H$ decays slower than $H_D$. For sufficiently large fluctuations \emph{above} the deterministic evolution, $H$ has a chance to ``freeze'' at a finite value rather than decrease to zero at late times. This is the main result of Ref.~\cite{deCesare:2016dnp}. The converse is also true. If fluctuations drive $H < H_D$, this will induce a slower decay of the energy density which in turn forces a faster $H \to 0$. Looking at specific realizations gives an idea of the possible outcomes, but does not give information of how likely each of them are. A proper analysis is required.

\subsection{Analysis of the Friedmann-Langevin equations}

We now proceed to solve these equations with a perturbative approach valid at early times. Our goal is to have some analytic control to estimate the statistical properties of the solutions of Eqs.~\eqref{langevin-cosmo}. We will validate this approach later by means of numerical simulations. First, let us consider the deterministic solutions of Eqs.~\eqref{langevin-cosmo} in the absence of noise,
\begin{equation} \label{deterministic-sol}
    \rho_D(t) = \frac{1}{6\pi \bar{G} t^2} ,\qquad H_D(t) = \frac{2}{3t},
\end{equation}
which correspond to the typical matter-dominated cosmological evolution. We use these to parametrize the stochastic solutions in the presence of noise as follows:
\begin{equation}
    \rho = \rho_D \left( 1+x \right) , \qquad H = H_D \left( 1+y \right),
\end{equation}
where $x \equiv \delta\rho/\rho_D$ and $y \equiv \delta H/H_D$ are the relative deviations from the deterministic solutions for the energy density and Hubble parameter respectively. When inserted back into Eqs. \eqref{langevin-cosmo}, and using the deterministic equations, we arrive at the following equivalent Langevin equations for the relative deviations:
\begin{subequations}
\begin{align}
 \dot{x} &= - 3 H_D \, y (1+x), \\
 \dot{y} &= \frac{\dot{H_D}}{H_D} \left[ x-y+ (1+x)\sigma \xi \right]. 
\end{align}
\end{subequations}
Using the explicit form of $H_D(t)$ and reparametrizing the time variable as $u = \log(t/t_i)$, these can be recast as
\begin{subequations} \label{langevin-cosmo-reparam}
\begin{align}
 \frac{dx}{du} &= -2 y(1+x), \\
 \frac{dy}{du} &= -x+y-(1+x)\sigma \xi,
\end{align}
\end{subequations}
which until now are fully equivalent to the original Eqs.~\eqref{langevin-cosmo}, i.e. nonlinear, but we have factored out the inconvenient power-law decay of the deterministic solutions.

At early times we can assume that the stochastic solutions do not deviate too much from the deterministic ones of Eq. \eqref{deterministic-sol}, i.e., $x, y \ll 1$. Expanding Eqs. \eqref{langevin-cosmo-reparam} linearly in this regime gives
\begin{subequations} \label{langevin-cosmo-reparam-linear}
\begin{align}
 \frac{dx}{du} &= -2 y, \\
 \frac{dy}{du} &= -x+y-\sigma \xi. 
\end{align}
\end{subequations}
 In terms of the original variables, the solutions to these
 perturbative equations read
\begin{subequations} \label{linear-sol}
\begin{align}
 \delta \rho &\equiv \rho_D \, x = \frac{2\rho_D}{3} \left[ x(t_i) \left(\frac{t_i}{t} + \frac{t^2}{2t_i^2} \right) + y(t_i) \left(\frac{t_i}{t} - \frac{t^2}{t_i^2} \right) - \frac{\sigma}{t} \int_{t_i}^{t} dt' \, \xi(t') + \sigma \, t^2 \int_{t_i}^{t} \frac{dt'}{t'^3} \, \xi(t') \right], \label{linear-drho} \\ 
 \delta H &\equiv H_D \, y = \frac{H_D}{3} \left[ x(t_i) \left(\frac{t_i}{t} - \frac{t^2}{t_i^2} \right) + y(t_i) \left(\frac{t_i}{t} + \frac{2t^2}{t_i^2} \right) - \frac{\sigma}{t} \int_{t_i}^{t} dt' \, \xi(t') - 2 \sigma \, t^2 \int_{t_i}^{t} \frac{dt'}{t'^3} \, \xi(t') \right]. \label{linear-dH}
\end{align}
\end{subequations}
In these perturbative solutions the stochastic contribution always enters linearly in the noise $\xi(t)$, which has a cumulative effect as it appears inside integrals, weighted against different power-law functions (constant and $t'^{-3}$ in this case). 

In solving the Friedmann-Langevin equations perturbatively we have also allowed for initial nonvanishing values for the perturbations which give a power-law behavior. Under the consistency relation of Eq.~\eqref{IC-consistency}, these nonvanishing initial conditions actually require initial values for the vacuum energy $\lambda_i \neq 0$ and/or a gravitational coupling $G_i \neq \bar{G}$, so one needs to be careful not to reintroduce the cosmological constant here. Our purpose in keeping these initial conditions for the fluctuations generic is to endow them with a stochastic origin, which we will discuss now.

Given the stochastic nature of the gravitational coupling we are considering in this work, it is reasonable to expect that even at time $t_i$ its value will stochastically differ from its mean, that is $G_i = \bar{G}(1+\sigma \xi_i)$. Then, again from Eq. \eqref{IC-consistency} this implies the consistency relation between the initial conditions is not sharp, but rather stochastic. In other words, one cannot initially have both a sharp value for the energy density and the Hubble parameter simultaneously, since both are related by $G_i$, which is in itself stochastic. Here we take the viewpoint that $\rho_i = \rho_D(t_i)$ takes an initially sharp value, since it is related to the matter content, and therefore it is the Hubble parameter which instead is initially stochastic,
\begin{equation}
H_i^2 =  \frac{8\pi}{3} \bar{G} (1+\sigma \xi_i) \rho_i.
\end{equation}
This is however just a convention and there is no loss of generality, as there is always a free parameter. 
In terms of the relative deviations $x$ and $y$, the above means
\begin{subequations} \label{stochastic-IC}
\begin{align}
 x(t_i) &= 0 ,\\
 y(t_i) &= \sqrt{1+\sigma \xi_i} - 1 \simeq \frac{1}{2} \sigma \xi_i - \frac{1}{8} \sigma^2 \xi_i^2 + \dots, \label{stochastic-yI}
\end{align}
\end{subequations}
 where the Taylor expansion is just indicative of one way to estimate the statistical properties of $y(t_i)$ in terms of those of $\xi_i$. Notice that the presence of the square root above signals that for a normally distributed $\xi_i$, $y(t_i)$ is ill defined. This is just the statement that, however unlikely, $G_i < 0$ values are possible. One way out is to assume that the initial step $\xi_i$ follows a half-normal distribution constraining it to be positive\footnote{This does not imply the positivity of subsequent stochastic steps $\xi(t)$ for $t > t_i$. Instead $\xi(t)$ follows a normal distribution. Then, for sufficiently small $\sigma$ one can make the probability of having $G(t) < 0$ at a given time arbitrarily small.}, i.e. $\xi_i > 0$. For a discussion regarding this point see Ref.~\cite{deCesare:2016dnp}. With this choice, the statistics of $\xi_i$ can be given in terms of a single dimensionful quantity, which we can express in terms of the initial value of the deterministic Hubble parameter $H_D^{(i)}$ and a dimensionless coefficient $\alpha$,
\begin{equation}
\langle \xi_i^n \rangle = \left( 2 \alpha H_D^{(i)} \right)^{n/2} \frac{\Gamma\left( \frac{n+1}{2} \right)}{\Gamma\left( \frac{1}{2} \right)}.
\end{equation}
The need to introduce a new parameter $\alpha H_D^{(i)}$ is rooted in the fact that the above quantities are statistical correlators of the noise at coincident times $t = t_i$, which may be divergent, as is the case for a white-noise correlation function, $\langle \xi(t) \xi(t) \rangle = R(0) = \delta(0).$  This requires some kind of regularization procedure to give meaning to such correlators. When taking an effective field theory point of view, where the noise is originated by integrating out degrees of freedom, it is natural to choose a cutoff for the noise correlators associated to the highest energy to which the macroscopic system is sensitive. In the current setup, the reference energy scale for the system is $H_D^{(i)} \equiv H_D(t_i) = \frac{2}{3t_i}$, which suggests we take $\alpha = 1$. Here however we choose to remain agnostic and instead vary $\alpha$ around this value in order to assess how the results depend on the choice of regulator. As we will see later, there is indeed a strong dependence.

The immediate consequence of the nonlinear relation between $y(t_i)$ and $\xi_i$ in Eq.~\eqref{stochastic-yI}, which in itself follows a half-normal distribution, is that the mean values of the fluctuations in Eq.~\eqref{linear-sol} do not both vanish,
\begin{equation} \label{mean-fluct}
    \langle \delta \rho \rangle = 0, \qquad \langle \delta H \rangle = \frac{H_D}{3} \langle y(t_i) \rangle \left(\frac{t_i}{t} + \frac{2t^2}{t_i^2} \right),
\end{equation}
where we used Eq.~\eqref{mean-noise}, or in other words the mean value of the Hubble parameter is not given by its deterministic counterpart of Eq.~\eqref{deterministic-sol}, due to the influence of the non-normally distributed initial conditions. Next we can compute the variances and the cross-correlations from Eqs. \eqref{linear-sol} by using $\langle \xi(t) \xi(t') \rangle = R(t-t')$ and making the assumption that the initial value of the noise $\xi_i$ and the noise at a later time $\xi(t)$ are uncorrelated, i.e. $\langle \xi_i^n \xi(t) \rangle = 0$, or equivalently 
$\langle y(t_i) \xi(t) \rangle = 0$.

Let us consider for example the variance of the Hubble parameter,
\begin{eqnarray}
 \text{Var}(H) = \langle \delta H^2 \rangle - \langle \delta H \rangle^2,
\end{eqnarray}
where we used Eq.~\eqref{mean-fluct}. Then, from Eqs.~\eqref{linear-dH} and \eqref{mean-fluct} we obtain
\begin{eqnarray} \label{var-H}
    \text{Var}(H) &=& \frac{H_D^2}{9} \Biggl[ \text{Var}(y(t_i)) \left(\frac{t_i}{t} + \frac{2t^2}{t_i^2} \right)^2 + \frac{\sigma^2}{t^2} \int_{t_i}^{t} dt' \int_{t_i}^{t} dt'' \left( 1 + \frac{2 t^3}{t'^3} \right)\left( 1 + \frac{2 t^3}{t''^3} \right) R(t'-t'') \Biggr]. \notag \\
\end{eqnarray} 
Note that the first term comes from the stochastic initial condition for $H$ and it is nonvanishing at $t = t_i$.

In order to make further progress either analytically or numerically it is necessary to assume some form for the noise correlation function $R(t-t')$. Let us consider a white-noise correlation function $R(t-t') = \delta(t-t')$. Immediately we can perform both integrations in Eq.~\eqref{var-H},
\begin{eqnarray} \label{var-H-white}
    \text{Var}(H) &=&  \frac{4}{81}\Biggl[\text{Var}(y(t_i)) \left( \frac{t_i^2}{4t^4} + \frac{1}{t\, t_i} + \frac{t^2}{t_i^4} \right) + \frac{\sigma^2}{t^4} (t - t_i) + \frac{4\sigma^2}{t} \left( \frac{1}{2t_i^2} - \frac{1}{2t^2} \right) \notag\\
    &&+ 4\sigma^2 t^2 \left( \frac{1}{5t_i^5} - \frac{1}{5t^5} \right) \Biggr] 
    \simeq \frac{4}{81} \left( \frac{\text{Var}(y(t_i))}{t_i^4} + \frac{4\sigma^2}{5t_i^5} \right) t^2 .
\end{eqnarray}
While most terms decay, we can see that there are terms that are actually growing as $\sim t^2$ [or $t^4$ relative to $H_D(t)^2$]. These results can be trusted up to the time $t \sim T_{\text{NP}}$ at which $\delta H$ becomes comparable with $H_D$, a condition that we can estimate by
\begin{equation}
    \langle \delta H \rangle + \text{Var}(\delta H)^{1/2} \sim H_D,
\end{equation}
which gives 
\begin{equation} \label{t_NP-cosmo}
    H_D^{(i)} \, T_{\text{NP}} \simeq \left[ 2\langle y(t_i) \rangle + \left( \text{Var}(y(t_i)) + \frac{6}{5} \sigma^2 H_D^{(i)} \right)^{1/2} \right]^{-1/2} \sim 
    \begin{cases}
        \sigma^{-1/2} (H_D^{(i)})^{-1/4} \quad & \alpha \lesssim 1, \\
        \sigma^{-1/2} (\alpha H_D^{(i)})^{-1/4} \quad & \alpha \gg 1.
    \end{cases}
\end{equation}
In the last expression we have estimated $\langle y(t_i)^n \rangle \sim \sigma^n (\alpha H_D^{(i)})^{n/2}$ and dropped $\mathcal{O}(1)$ factors. Notice that $T_{\text{NP}}$ is the characteristic time at which a typical stochastic realization departs from the deterministic evolution $H_D(t)$ by an $\mathcal{O}(1)$ factor, a measure of when the stochastic effects become dominant.

\begin{figure}
    \centering
    \includegraphics[width=0.49\textwidth]{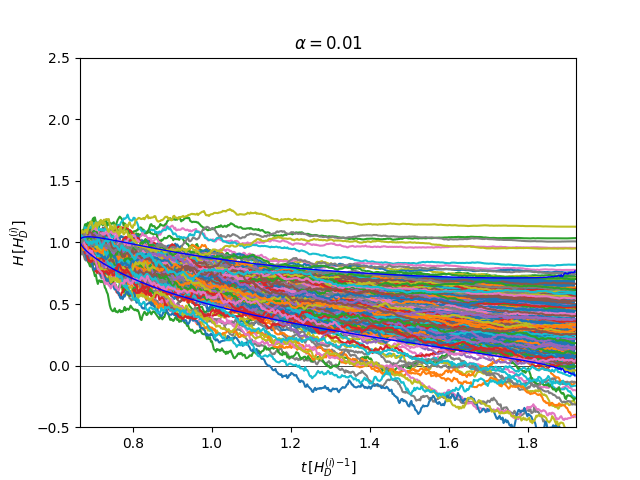}
    \includegraphics[width=0.49\textwidth]{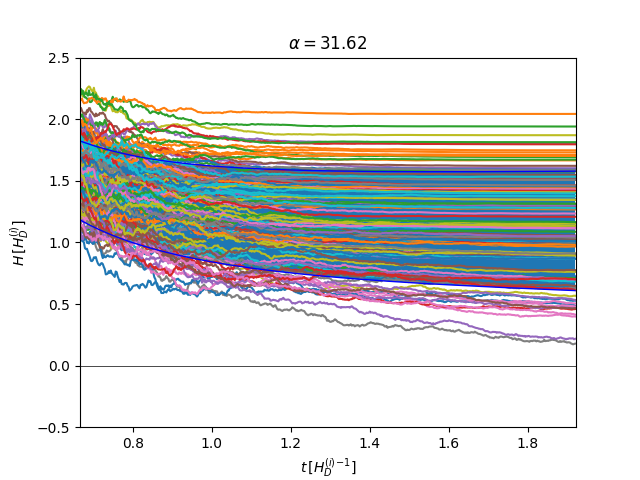}
    \includegraphics[width=0.49\textwidth]{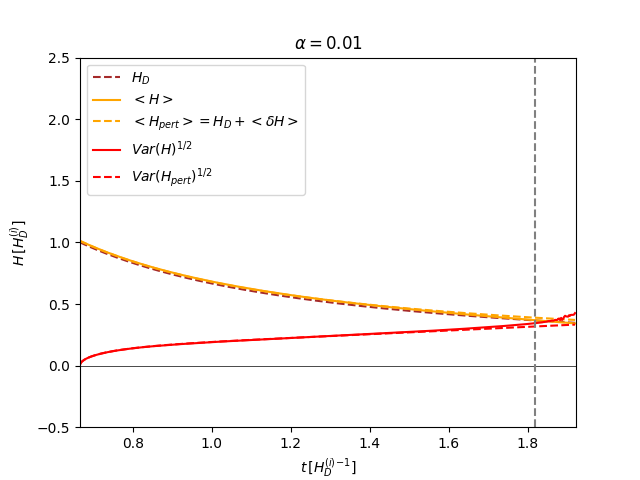}
    \includegraphics[width=0.49\textwidth]{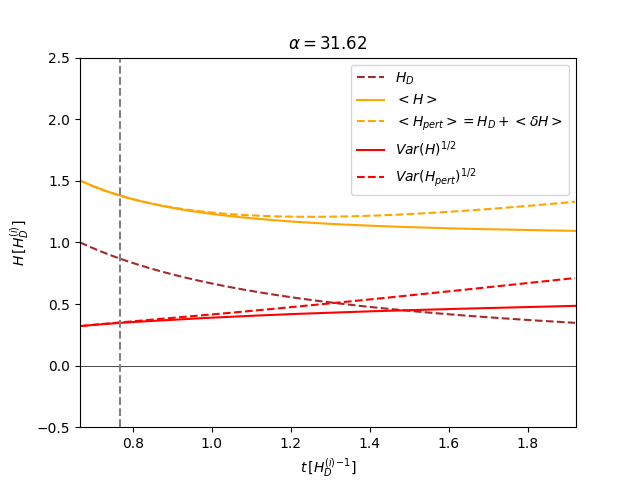}
    \caption{Top panels show a subsample of $100$ numerical realizations of the stochastic process with parameters $\sigma^2 = 0.1 \, (H_D^{(i)})^{-1}$, $\bar{G}=1$. These are nonrealistic values and were chosen as in Ref.~\cite{deCesare:2016dnp} for the sake of comparison. Only the Hubble parameter $H$ is shown. The blue envelope encloses a region of width $2\times \sqrt{\text{Var}(H)}$ around the mean value $\langle H \rangle$. Bottom panels show a comparison of the perturbative expectations for the stochastic mean value (dashed orange) and variance (dashed red) of the Hubble parameter $H$, against those computed from the full set of $N=10^5$ numerical realizations (solid orange and red respectively). The deterministic value for the Hubble parameter $H_D(t)$ for a matter-dominated universe is also shown (dashed brown) for reference. Good agreement can be seen within the regime of validity of the perturbative approximation (the vertical dashed gray line indicates the time $T_{\text{NP}}$). White noise is assumed in all cases. The left panels assume a small value for the regulator $\alpha H_D^{(i)}$, namely $\alpha = 0.01$, while the right panels instead assume a large value $\alpha = 31.62$. A strong dependence on this value can be observed, as it dictates how much the initial values are spread towards larger values of $H_i$ (a positive half-normal distribution for the initial noise $\xi_i$ is assumed). For a small spread (left panels), the time evolution of a given realization for $H$ can either go to a constant or to negative values with similar probability. This is reflected on the mean value $\langle H \rangle$ very closely following the deterministic prediction $H_D(t)$. In contrast, for large spread (right panels) there is a larger probability of the realizations for $H$ staying constant, which is also observed on the mean value $\langle H \rangle$ decaying much more slowly than $H_D(t)$.}
    \label{fig:plot-cosmo}
\end{figure}

At this point we expect the nonlinearities to either curb or enhance this growth. In Fig.~\ref{fig:plot-cosmo} we show  the time evolution of the mean value and variance of the Hubble parameter $H$, both from our perturbative approximation (dashed) and numerical simulations (solid), for two very different choices of $\alpha$. While good agreement is observed within the regime of validity of the perturbative treatment, i.e., Eq.~\eqref{t_NP-cosmo}, this can be very short for larger values of $\alpha$. More importantly, there is an observed change in behavior as $\alpha$ varies. For small values of $\alpha$, the stochastic initial conditions have little spread and therefore the different realizations exhibit varied behaviors over time. In this case, it is as likely for a given realization to plunge to $H < 0$ values as it is for it to stay at roughly constant ones. This changes dramatically as $\alpha$ increases, especially since we are taking the initial noise $\xi_i$ to follow a positive half-normal distribution. This leads to a wide spread of initial conditions biased towards larger values of $H_i$ compared to $H_D^{(i)}$, leading to many more solutions that resemble a cosmological constant than decaying ones.

To summarize, the analysis presented here shows that stochastic fluctuations of $G$ may produce a cosmological evolution such that the Hubble parameter tends to a positive constant at long times, in agreement with Ref.~\cite{deCesare:2016dnp}. This, however, is not a robust prediction of the model as it strongly depends on the assumption of stochastic, biased initial conditions, in particular on the value of the regulator $\alpha H_D^{(i)}$.

\section{Stochastic effects on binary systems} \label{sec:binary}

Observational data from binary systems, and in particular from LLR,  provide stringent bounds on modifications to General Relativity, including tests of the equivalence principle, the inverse square Newton law, and eventual variations of the strength of gravity \cite{Muller:2007zzb,Merkowitz:2010kka}.
For example, assuming a time dependence of the gravitational constant
\begin{equation}\label{time-dep-G}
    G(t)=G_i\left(1 + \frac{\dot G}{G}(t_i)(t-t_i)+ \frac{1}{2}\frac{\ddot G}{G}(t_i)(t-t_i)^2\right)\, ,
\end{equation}
the whole set of data of LLR implies the following limits \cite{Biskupek:2020fem}
\begin{eqnarray}
   && \frac{\dot G}{G}=(-5.0\pm 9.6)\times 10^{-15}yr^{-1} \, ,\nonumber\\
    && \frac{\ddot G}{G}=(1.6\pm 2.0)\times 10^{-16}yr^{-2}\, .
\end{eqnarray}

These values are obtained after a sophisticated analysis that takes into account the effects of ocean tidal bulges, the precession of the lunar orbit's plane, etc., but the main idea is elementary: 
the time dependence of the gravitational coupling will induce changes in the radial size of the orbit $r$ and in the period of revolution $P$ \cite{Nordtvedt1999}:
\begin{equation}
\frac{\dot r}{r}=-\frac{\dot G}{G}\quad\quad \quad
\frac{\dot P}{P}=-2\frac{\dot G}{G}\,\, ,
\end{equation}
so that the Moon-Earth distance should change linearly with time
\begin{equation}
r(t)\simeq r(t_i) \left(1 -\frac{\dot G}{G}(t_i)(t-t_i)\right)\, .
\end{equation}

We will now discuss, also at an elementary level, the effects of a stochastic variation of $G$ on the distance between astronomical bodies in a Newtonian binary system. We consider the effective one-body problem of a binary system in Newtonian gravity. We can always reduce this to an effective equation of motion for the radial coordinate $r$ by using conservation of angular momentum,
\begin{equation}
    \frac{d^2 r}{dt^2} = - \frac{G(t)M}{r^2} + \frac{l^2}{r^3},
\end{equation}
where $M$ is the total mass of the binary system and $l$ the orbital angular momentum per unit reduced mass. Here we allow once again for a time dependence of the gravitational coupling $G$ of stochastic origin, given by Eq.~\eqref{stochastic-G},  with noise statistics given by Eqs.~\eqref{mean-noise} and \eqref{correlation}.

We consider for simplicity quasicircular motion and expand $r = \bar{r}_0 + \delta r$, around the deterministic circular motion radius $\bar{r}_0 = l^2/\bar{G}M$ ($\ddot{\bar{r}}_0 = 0$). The perturbation $\delta r \ll \bar{r}_0$ satisfies the following equation up to second order, 
\begin{equation} \label{orbital-pert}
    \delta\ddot{r} + \bar{\Omega}^2_0 \left( 1 - 2\sigma \xi(t) \right) \delta r = - \bar{r}_0 \bar{\Omega}^2_0 \, \sigma \xi(t) + 3 \bar{\Omega}^2_0 (1-\sigma \xi(t)) \frac{\delta r^2}{\bar{r}_0} + \mathcal{O}(\delta r^3),
\end{equation}
where $\dot{} \equiv \partial_t$ and $\bar{\Omega}^2_0 = \bar{G}M/\bar{r}_0^3$ is the squared orbital frequency of the circular motion. In what follows we will study the linearized equation, but here we have kept the quadratic terms in $\delta r$ to emphasize that, having a positive sign on the right-hand side of Eq.~\eqref{orbital-pert} they will tend to push $\delta r$ towards positive values when they become relevant. Dropping these for now, we get for the linearized equation a harmonic oscillator
\begin{equation} \label{orbital-pert-linear}
    \delta\ddot{r} + \Omega^2(t) \delta r = f(t),
\end{equation}
with time-dependent stochastic squared frequency
\begin{equation}
 \Omega^2(t) = \bar{\Omega}^2_0 \left( 1 - 2\sigma \xi(t) \right),
\end{equation}
and a stochastic additive noise
\begin{equation}
 f(t) = - \bar{r}_0 \bar{\Omega}^2_0 \, \sigma \xi(t).
\end{equation}

Similar stochastic equations have been studied thoroughly in the literature, in particular, results are well known when either one of the additive or multiplicative noises are present separately. 

The case of multiplicative noise has been studied in different contexts, in particular in the analysis of wave propagation in random media \cite{papanicolau1971stochastic}. The solutions of the stochastic equation Eq.~\eqref{orbital-pert-linear} with $f(t) = 0$ show an exponential increase 
\begin{equation} \label{multiplicative-variance}
\langle \delta r^2\rangle_{\text{multipl}} \propto e^{4\sigma^2 \bar{\Omega}_0^2 Re[S(2 \bar{\Omega}_0)]t}\, ,
\end{equation}
where $S(\omega)$ is the Fourier transform of the noise correlation function, Eq.~\eqref{correlation},
\begin{equation} \label{Fourier-correlation}
S(\omega)=\int_0^\infty dt\, R(t) e ^{i\omega t} \, .
\end{equation}
The exponential growth is the stochastic counterpart of the well-known phenomenon of parametric resonance.

The effect of the additive noise is more closely related to the usual Brownian motion, and the growth of $\langle \delta r^2\rangle $
is not exponential but polynomial. For instance, assuming a Gaussian colored noise with correlation function
\begin{equation}\label{colored-noise}
R(t-t')=\frac{1}{\tau_c}e^{-\vert t-t'\vert/\tau_c}\, ,
\end{equation}
in the large time limit the variance is given by \cite{PhysRevE.48.4309}
\begin{equation}\label{additive-variance}
\langle \delta r^2\rangle_{\text{add}} = \frac{\bar r_0^2 \Omega_0^2 \sigma^2}{(\tau_c^2\Omega_0^2+1)}t\, .
\end{equation}

These two simplified cases with only a single type of noise active at a time give us a hint of what to expect when both effects are combined. On the one hand, the additive noise ensures that any system will be kicked out of the equilibrium and start to oscillate. The multiplicative noise will then, over a longer timescale, induce an exponential growth via stochastic parametric resonance. However intuitive this picture may be, these effects are not so easily combined analytically. In order to take them both into account simultaneously in a rigorous way, we need to dive into Eq.~\eqref{orbital-pert-linear} with MSA methods \cite{bender78:AMM,papanicolau1971stochastic}. In the next subsection we will show how these methods can be used to derive stochastic mean values and variances in such a scenario. Readers not interested on the technical details can skip ahead to Sec.~\ref{sec:constraints}, where the results are discussed in the light of LLR observational constraints.

\subsection{Stochastic multiple scale analysis} \label{sec:MSA}

The basis of the MSA is the understanding that solutions to equations like \eqref{orbital-pert-linear} exhibit distinct behaviors at different time scales. As already shown by the known result in Eq.~\eqref{multiplicative-variance} for multiplicative noise, we expect an exponential growth that onsets on a timescale $T_l \sim (\sigma \bar{\Omega}_0)^{-2}$ which is much longer than the period of oscillation $T_s = 2\pi \, \bar{\Omega}_0^{-1}$, provided $\sigma^2 \bar{\Omega}_0 \ll 1$. With this in mind, we define the expansion parameter $\epsilon^2 \equiv T_s/(2\pi \, T_l) = \sigma^2 \bar{\Omega}_0$, and a new \emph{slow} time variable $\tau = \epsilon^2 t$. 

We start our discussion by rewriting Eq.~\eqref{orbital-pert-linear} in terms of the dimensionless variable $z \equiv \delta r/\bar{r}_0$, replacing also $\sigma$ in favor of $\epsilon$, and rescaling the noise to be dimensionless as well, i.e. $\epsilon \eta(t) = -2 \sigma \xi(t)$,
\begin{equation} \label{MSA-eq}
    \ddot{z} + \omega^2 \left( 1 + \epsilon \eta(t) \right) z = - \omega^2 \gamma \epsilon \, \eta(t).
\end{equation}
Here we have also introduced $\gamma = -1/2$ as a bookkeeping parameter and wrote $\omega \equiv \bar{\Omega}_0$ to alleviate the notation. Expanding a solution $z(t)$ of Eq.~\eqref{MSA-eq} in powers of $\epsilon$ as follows:
\begin{equation} \label{z-pert}
    z(t) = Z_0(t,\tau) + \epsilon Z_1(t,\tau) + \epsilon^2 Z_2(t,\tau) + O(\epsilon^3),
\end{equation}
and then, introducing this expansion in \eqref{MSA-eq} we get the set of equations
\begin{equation}
\begin{cases} 
\label{MSA-eqs-pert}
    \partial_t^2 Z_0 + \omega^2 Z_0 = 0, \\
    \partial_t^2 Z_1 + \omega^2 Z_1 + \omega^2 \eta(t) \left( Z_0 + \gamma \right) = 0, \\
    \partial_t^2 Z_2 + \omega^2 Z_2 + \omega^2 \eta(t) Z_1 + 2 \partial_{\tau t}^2 Z_0 = 0,
\end{cases}
\end{equation}
where we are treating $\tau$ and $t$ as independent variables. If one ignores the dependence on the \emph{slow} time $\tau$, Eqs.~\eqref{z-pert} and \eqref{MSA-eqs-pert} are just a standard perturbative expansion in $\epsilon$. However, here the approach will be to first solve for the \emph{fast} time $t$, which expectedly will give rise to secularly growing contributions for stochastic mean values. On a second stage, these contributions are resummed by the $\tau$ dependence of the otherwise constant coefficients coming out of the first stage.

At zeroth order the solution $Z_0$ to the first equation in \eqref{MSA-eqs-pert} is just that of a harmonic oscillator with frequency $\omega$,
\begin{equation} \label{Z0}
    Z_0(t,\tau) = A(\tau) e^{i\omega t} + B(\tau) e^{-i\omega t},
\end{equation}
where $A(\tau)$ and $B(\tau)$ are to be determined later on. Moving on to the first order, the harmonic oscillator is now subject to a source term $-\omega^2 \eta(t) \left( Z_0 + \gamma \right)$, and therefore we can formally solve for $Z_1$ by using the corresponding retarded Green's function
\begin{equation} \label{Z1}
    Z_1(t,\tau) = C(\tau) e^{i\omega t} + D(\tau) e^{-i\omega t} - \omega \int_0^t dt' \sin[\omega (t-t')] \eta(t') \left[ A(\tau) e^{i\omega t'} + B(\tau) e^{-i\omega t'} + \gamma \right]\, .
\end{equation}
Here $C(\tau)$ and $D(\tau)$ are arbitrary functions, fixed by the initial conditions, that can be chosen to vanish without loss of generality (i.e. imposing the initial conditions on $Z_0$).  In the above equation, 
the integration on $t'$ does not affect the $\tau$ dependence, as they are independent variables. At this point we can compute the mean value of $Z_1$ and trivially obtain 
$\mv{Z_1(t,\tau)} = 0$, 
which implies the absence of secular terms at this order. It is then necessary to go to higher order to find a condition on the coefficients $A(\tau)$ and $B(\tau)$. Solving similarly for $Z_2$ we get
\begin{equation}\label{eq:Z2}
\begin{split}
    Z_2(t,\tau) = & - 2 i \int_0^t dt' \sin[\omega(t-t')] \left[ A'(\tau) e^{i\omega t'} - B'(\tau) e^{-i\omega t'} \right] \\
    & +\omega^2 \int_0^t dt' \sin[\omega(t-t')] \eta(t') \bigg\{  \int_0^{t'} dt'' \sin[\omega (t'-t'')] \eta(t'') \left[ A(\tau) e^{i\omega t''} + B(\tau) e^{-i\omega t''} + \gamma \right] \bigg\} ,
\end{split}
\end{equation}
and therefore
\begin{equation} \label{Z2--mean}
\begin{split}
    \mv{Z_2(t,\tau)} = & \int_0^t dt' \sin[\omega(t-t')] \bigg\{ -2 i \left[ A'(\tau) e^{i\omega t'} - B'(\tau) e^{-i\omega t'} \right] \\
    & + 4\omega \int_0^{t'} dt'' \sin[\omega (t'-t'')] R(t'-t'') \left[ A(\tau) e^{i\omega t''} + B(\tau) e^{-i\omega t''} + \gamma \right] \bigg\}.
\end{split}
\end{equation}

The coefficients $A(\tau)$ and 
$B(\tau)$ are fixed in such a way that there are no secular terms in the mean value $\langle Z_2(t,\tau)\rangle$ (this procedure reproduces the results of Ref.~\cite{papanicolau1971stochastic} for the case of multiplicative noise, as described in Ref.~\cite{Mantinan:2022vcn}). Note that secular terms do arise when there are contributions proportional to $e^{\pm i \omega t'}$ inside the curly brackets under the integral over $t'$. As shown in the Appendix, the coefficients $A(\tau)$ and 
$B(\tau)$ that avoid such contributions are 
\begin{equation} \label{A-B-mean0}
\begin{cases} 
    A(\tau) = A(0) \, e^{\left[S^*(2\omega) - S(0) \right] \omega \epsilon^2 t},  \\
    B(\tau) = B(0) \, e^{\left[S(2\omega) - S(0) \right] \omega \epsilon^2 t}   ,
\end{cases}
\end{equation}
where we have restored $t$ in favor of $\tau$. Plugging these results into the expansion \eqref{z-pert} we obtain
\begin{equation}
    z(t) = e^{\left[ \text{Re}[S(2{\omega})] - S(0) \right] \omega \epsilon^2 t} 
   \left[ A(0) \, e^{i (1 - \epsilon^2 \text{Im}[S(2\omega)]) \omega t} + B(0) \, e^{-i (1 - \epsilon^2 \text{Im}[S(2\omega)]) \omega t} \right] + O(\epsilon).
\end{equation}
An important feature of this result is that, although only valid at leading order in $\epsilon$, since the $O(\epsilon)$ terms are linear in the noise, the mean value actually enjoys next-to-leading order accuracy,
\begin{equation}
   \langle z(t)\rangle = e^{\left[ \text{Re}[S(2{\omega})] - S(0) \right] \omega \epsilon^2 t} 
   \left[ A(0) \, e^{i (1 - \epsilon^2 \text{Im}[S(2\omega)]) \omega t} + B(0) \, e^{-i (1 - \epsilon^2 \text{Im}[S(2\omega)]) \omega t} \right] + O(\epsilon^2) .
\end{equation}

With this result, we see that the stochastic mean value of the evolution of linear perturbation, $z(t)$, generically receive corrections with respect to its deterministic counterpart in the form of an exponential factor and a shift in the frequency. However, these effects can be important or not depending on the noise correlation function $R(t-t')$. For example, for a white-noise correlation function we have that $S(\omega)=1/2$, and therefore we find that both these effects are not present at the level of the mean value in our MSA approach. It is also worth noting that, at this level, there is no effect of the additive noise, as the absence of our bookkeeping parameter $\gamma$ evidences. This was to be expected considering the vanishing mean of the noise, Eq.~\eqref{mean-noise}. To properly assess the effect of the noise we need to also compute the variance.

We now proceed to discuss the stochastic average of the perturbation squared, $\langle z(t)^2 \rangle$, needed for the computation of the variance. Here the procedure is analogous to what we have just done for the mean value. We compute the stochastic averages up to second order in $\epsilon$, search for the secular terms and fix the dependence of zeroth order coefficients with the \emph{slow} time $\tau$ in order to cancel them. Importantly, these will be different from the ones found above, Eq.~\eqref{A-B-mean0}, as the secular terms we need to resum here will be different than those in $\langle z(t) \rangle$.

Using the same expansion as before, Eq.~\eqref{z-pert}, we now have
\begin{equation}
    z(t)^2 = Z_0(t,\tau)^2 + 2 \epsilon Z_0(t,\tau) Z_1(t,\tau) + \epsilon^2 \left( Z_1(t,\tau)^2 + 2 Z_0(t,\tau) Z_2(t,\tau) \right) + \mathcal{O}(\epsilon^3),
\end{equation}
and therefore
\begin{equation}
    \langle z(t)^2 \rangle = Z_0(t,\tau)^2 + 2 \epsilon Z_0(t,\tau) \langle Z_1(t,\tau) \rangle + \epsilon^2 \left[ \langle Z_1(t,\tau)^2 \rangle + 2 Z_0(t,\tau) \langle Z_2(t,\tau) \rangle \right] + \mathcal{O}(\epsilon^3),
\end{equation}
where we have used that $Z_0(t,\tau)$ is independent of the noise. As before, the term linear in $\epsilon$ becomes irrelevant in that it does not contain secular terms, while it also important that now the structure of the zeroth order part is 
\begin{equation} \label{Z0Z0}
    Z_0(t,\tau)^2 = A(\tau)^2 \, e^{2i\omega t} + 2 A(\tau) B(\tau) + B(\tau)^2 \, e^{-2i\omega t}.
\end{equation}
This means there are actually three independent coefficients to fix $A(\tau)^2$, $B(\tau)^2$, and $2 A(\tau) B(\tau)$, which is consistent within the stochastic MSA method.

The secular terms will be then found in the second order part
\begin{equation} \label{second-order-zz}
    \langle Z_1(t,\tau)^2 \rangle + 2 Z_0(t,\tau) \langle Z_2(t,\tau) \rangle,
\end{equation}
of which we already have a partial computation, i.e., the second term, by combining Eqs.~\eqref{Z0} and \eqref{Z2--mean}. Then, we only need to compute $\langle Z_1(t,\tau)^2 \rangle$. This is done in the Appendix, where we show that
\begin{eqnarray} \label{Z1Z1--mean2}
    \langle Z_1(t,\tau)^2 \rangle &=&  8 \omega \int_0^t dt' \sin[\omega (t-t')] \left[ A(\tau) e^{i\omega t'} + B(\tau) e^{-i\omega t'} + \gamma \right] \notag\\
    &&\times \int_0^{t'} dt'' \sin[\omega (t-t'')] R(t'-t'') \left[ A(\tau) e^{i\omega t''} + B(\tau) e^{-i\omega t''} + \gamma \right].    
\end{eqnarray}
Inserting Eqs.~\eqref{Z0}, \eqref{Z2--mean}, and \eqref{Z1Z1--mean2} into Eq.\eqref{second-order-zz} we obtain the second order part of $\langle z(t)^2\rangle$. The calculation of the coefficients that avoid the secular terms is described in the Appendix. They are given by 
\begin{equation} \label{AA-BB-mean}
\begin{cases} 
    [A(\tau)]^2 = A(0)^2 \, e^{2 \left[S^*(2\omega) - 2S(0) \right] \omega \epsilon^2 t},  \\
    [B(\tau)]^2 = B(0)^2 \, e^{2 \left[S(2\omega) - 2S(0) \right] \omega \epsilon^2 t} , \\
     [2 A(\tau) B(\tau)] = 2 A(0) B(0) \, e^{4 \omega \text{Re}[S(2\omega)] \epsilon^2 t} + \gamma^2 \, \frac{\text{Re}[S(\omega)]}{\text{Re}[S(2\omega)]} \left( e^{4 \omega \text{Re}[S(2\omega)] \epsilon^2 t} - 1 \right) \, .
     \end{cases}
\end{equation}
Plugging these results back into Eq.~\eqref{Z0Z0} we obtain the main result of this subsection
\begin{eqnarray} \label{z2-mean}
    \langle z(t)^2 \rangle &=& e^{2 \left[\text{Re}[S(2\omega)] - 2S(0) \right] \omega \epsilon^2 t} \left[ A(0)^2 \, e^{2i (1 - \epsilon^2 \text{Im}[S(2\omega)]) \omega t} + B(0)^2 \, e^{-2i (1 - \epsilon^2 \text{Im}[S(2\omega)]) \omega t} \right] \notag \\
    &&+ 2 A(0) B(0) \, e^{4 \omega \text{Re}[S(2\omega)] \epsilon^2 t} + \gamma^2 \, \frac{\text{Re}[S(\omega)]}{\text{Re}[S(2\omega)]} \left( e^{4 \omega \text{Re}[S(2\omega)] \epsilon^2 t} - 1 \right) \, .
\end{eqnarray}

As a partial check of our calculation, we can show that these results reproduce those associated with the presence of either multiplicative or additive noises in isolation. The former is obtained in the limit $\gamma\to 0$, and the results give the typical exponential growth of parametric stochastic resonance, [see Eq.\eqref{multiplicative-variance}]. The latter is obtained in the limit $\epsilon\to 0, \gamma\epsilon\to const$, as can be deduced from Eq.~\eqref{MSA-eq}. In this limit $\langle z(t)^2\rangle$ grows linearly in time, as in the usual Brownian motion. For the particular colored noise given in Eq.\eqref{colored-noise}, $\langle z(t)^2\rangle$ reproduces the result in Eq.\eqref{additive-variance}.
 
It is worth noting that the terms associated with multiplicative noise alone, i.e., first and second in Eq.~\eqref{z2-mean}, grow exponentially with a rate that is only sensitive to the Fourier transform of the noise correlation evaluated at two times the natural frequency $\omega$. This is typical of parametric resonance. However, these terms have an amplitude that is proportional to the initial conditions. In contrast, the third term is associated with the combination of both multiplicative and additive noises. It also exhibits an exponential growth with the same rate as the standard parametric resonance effect, but it starts at linear order in $t$ (due to the $-1$ piece) with an overall amplitude proportional to $\gamma^2$.  This is interesting because, in addition to being independent of the initial condition for the perturbation, it will be, in general larger in amplitude (see below).  In other words, the presence of additive noise ensures that deviations from 
$z=0$
will always exist, and that they will be exponentially enhanced by the parametric resonance effect induced by the multiplicative noise.
  
\subsection{Constraints on the amplitude of the noise from LLR} \label{sec:constraints}

We now describe a concrete application of our results to illustrate how observational data from LLR can provide a constraint on the amplitude of the stochastic fluctuations of the gravitational coupling. Indeed, the current agreement of said data with the classical prediction for the Earth-Moon motion implies that 
\begin{equation} \label{LLR-bound}
    \langle \left( r_{obs} - r_{cl} \right)^2 \rangle \leq \delta_{\text{LLR}}^2,
\end{equation}
where $r_{obs} = \bar{r}_0 + \delta r$ is the observed radial Earth-Moon distance while $r_{cl}$ is the classical expectation in the absence of noise and $\delta_{\text{LLR}}$ is the typical precision of the laser ranging measurements. 

We can express the above equivalently as
\begin{equation} \label{LLR-bound-2}
    \text{Var}(\delta r) + (\langle \delta r \rangle - \delta r_{cl})^2 \leq \delta_{\text{LLR}}^2,    
\end{equation}
with $\delta r_{cl} = r_{cl} - \bar{r}_0$. Using our previous results, valid in the linear regime of Eq.~\eqref{orbital-pert}, choosing as initial condition $A(0) = B(0) = (1/2)\delta r(0)/\bar{r}_0$, and reinstating the original parameters (including $\gamma = -1/2$), we obtain 
\begin{eqnarray}
    \langle \delta r (t) \rangle &=& \delta r(0) \, e^{\left[ \text{Re}[S(2\bar{\Omega}_0)] - S(0) \right] \bar{\Omega}_0^2 \sigma^2 t} \cos\left[ \bar{\Omega}_0 \left(1 - \bar{\Omega}_0 \sigma^2 \text{Im}[S(2\bar{\Omega}_0)]\right)t \right] + \mathcal{O}(\bar{\Omega}_0 \sigma^2), \label{pert1-mean} \\
    \langle \delta r(t)^2 \rangle &=& \frac{1}{2} \delta r(0)^2 \, e^{2 \left[ \text{Re}[S(2\bar{\Omega}_0)] - 2S(0) \right] \bar{\Omega}_0^2 \sigma^2 t} \cos\left[ 2 \bar{\Omega}_0 \left(1 - \bar{\Omega}_0 \sigma^2 \text{Im}[S(2\bar{\Omega}_0)]\right)t \right] \notag\\
    &&+ \frac{1}{2} \delta r(0)^2 \, e^{4 \text{Re}[S(2\bar{\Omega}_0)] \bar{\Omega}_0^2 \sigma^2 t} + \frac{\bar{r}_0^2}{4} \, \frac{\text{Re}[S(\bar{\Omega}_0)]}{\text{Re}[S(2\bar{\Omega}_0)]} \left( e^{4 \text{Re}[S(2\bar{\Omega}_0)] \bar{\Omega}_0^2 \sigma^2 t} - 1 \right) \notag \\
    &&+ \mathcal{O}(\bar{\Omega}_0 \sigma^2), \label{pert2-mean}
\end{eqnarray}
and
\begin{eqnarray} \label{pert-var}
    \text{Var}(\delta r(t)) &=& \frac{1}{2} \delta r(0)^2 \, e^{2\left[ \text{Re}[S(2\bar{\Omega}_0)] - S(0) \right] \bar{\Omega}_0^2 \sigma^2 t} \left[ e^{- 2S(0) \bar{\Omega}_0^2 \sigma^2 t} - 1 \right] \cos\left[ 2 \bar{\Omega}_0 \left(1 - \bar{\Omega}_0 \sigma^2 \text{Im}[S(2\bar{\Omega}_0)]\right)t \right] \notag\\
    &&+ \frac{1}{2} \delta r(0)^2 \, e^{4 \text{Re}[S(2\bar{\Omega}_0)] \bar{\Omega}_0^2 \sigma^2 t} \left[ 1 - e^{- 2\left[ \text{Re}[S(2\bar{\Omega}_0)] + S(0) \right] \bar{\Omega}_0^2 \sigma^2 t} \right] \notag\\
    &&+ \frac{\bar{r}_0^2}{4} \, \frac{\text{Re}[S(\bar{\Omega}_0)]}{\text{Re}[S(2\bar{\Omega}_0)]} \left( e^{4 \text{Re}[S(2\bar{\Omega}_0)] \bar{\Omega}_0^2 \sigma^2 t} - 1 \right) \notag \\
    &&+ \mathcal{O}(\bar{\Omega}_0 \sigma^2).
\end{eqnarray}
These results show that, for orbits with small eccentricity ($e \simeq \delta r(0)/\bar{r}_0 \ll 1$), the classical solution $\delta r_{cl}$ (obtained in the limit $\sigma\to 0$) is modified in different ways due to the presence of the noise. On the one hand, the multiplicative noise induces a shift in the frequency of the classical oscillations, proportional to the imaginary part of $S(2\bar\Omega_0)$. The variance of the amplitude grows exponentially with a rate proportional to the real part of $S(2\bar\Omega_0)$. On the other hand, the third terms in Eqs.~\eqref{pert2-mean} and \eqref{pert-var} represent a concurrent effect of both multiplicative and additive noise, and will be the dominant terms as long as $\delta r(0)\ll \bar r_0$ [if we assume that $\text{Re}[S(\bar{\Omega}_0)]/\text{Re}[S(2\bar{\Omega}_0)]= \mathcal{O}(1)$]. As expected, the noise becomes relevant when the typical  scale  of temporal stochastic variations is of order $\bar\Omega_0^{-1}$. 

We stress that the above results are derived in the linear approximation, assuming  $\delta r\ll r_0$. Therefore, in the exponential regime (when $ \text{Re}[S(2\bar{\Omega}_0)] \bar{\Omega}_0^2 \sigma^2 t \gtrsim 1$), the last term proportional to $\bar{r}_0^2$ would violate this assumption, unless the noise spectrum is sufficiently peaked around $2\bar\Omega_0$. This would not be a limitation when applying these results to linear systems, as those considered in Refs.~\cite{Parikh:2020fhy,Parikh:2020nrd}.

For the purpose of finding a bound on the amplitude of the noise $\sigma$, we consider again a white-noise correlation function $R(t-t') = \delta(t-t')$. Notice that in this case $\sigma$ drops from Eq.~\eqref{pert1-mean} (at leading order), and therefore $\langle \delta r \rangle = \delta r_{cl} + \mathcal{O}(\bar{\Omega}_0 \sigma^2)$. Moreover, our result Eq.~\eqref{pert-var} specializes to
\begin{eqnarray} \label{pert-var-white}
    \text{Var}(\delta r(t)) &=& \frac{1}{2} \delta r(0)^2 \left( e^{- \bar{\Omega}_0^2 \sigma^2 t} - 1 \right) \cos\left( 2 \bar{\Omega}_0 t \right) + \frac{1}{2} \delta r(0)^2 \, e^{2 \bar{\Omega}_0^2 \sigma^2 t} \left( 1 - e^{- 2 \bar{\Omega}_0^2 \sigma^2 t} \right) \notag\\
    &&+ \frac{\bar{r}_0^2}{4}  \left( e^{2 \bar{\Omega}_0^2 \sigma^2 t} - 1 \right) + \mathcal{O}(\bar{\Omega}_0 \sigma^2).
\end{eqnarray}
At short times, $\bar\Omega_0^2 \sigma^2 t\ll 1$, the bound expressed in Eqs.~\eqref{LLR-bound} and \eqref{LLR-bound-2} can be written as
\begin{eqnarray}\label{bound LLR fin}
    \left[\text{Var}(\delta r(t))\right]^{1/2} &\simeq& \left[ 1 + e^2 \left( 2 - \cos\left( 2 \bar{\Omega}_0 t \right) \right)  \right]^{1/2} \, \bar{r}_0 \bar{\Omega}_0 \sigma \sqrt{\frac{t}{2}} \leq \delta_{\text{LLR}},
\end{eqnarray}
where $e \simeq \delta r(0)/\bar{r}_0$ is the eccentricity of the orbit, which in the case of the Moon is $e \simeq 5 \%$, and therefore for our purposes of giving an order of magnitude estimate we can drop the second term in the brackets. Considering that the LLR experiment has achieved a few-millimeter range precision \cite{Murphy_2012}, and assuming fifteen years of observation with that precision, Eq.\eqref{bound LLR fin} translates into a bound in the amplitude of the noise of order
\begin{equation} \label{LLR-bound-3}
    \sigma_{\text{LLR}} \lesssim 10^{-13} \, \text{yr}^{1/2}.
\end{equation}
This is a very small value that justifies the short-times assumption above.

\begin{figure}[t!] 
\begin{center} 
\includegraphics[width=0.49\linewidth]{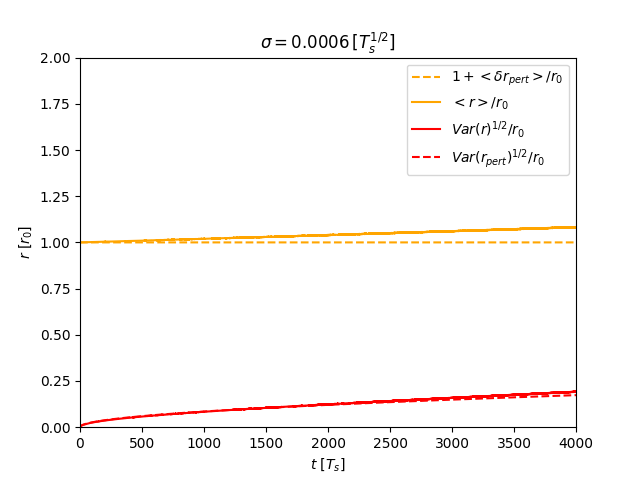}
\includegraphics[width=0.49\linewidth]{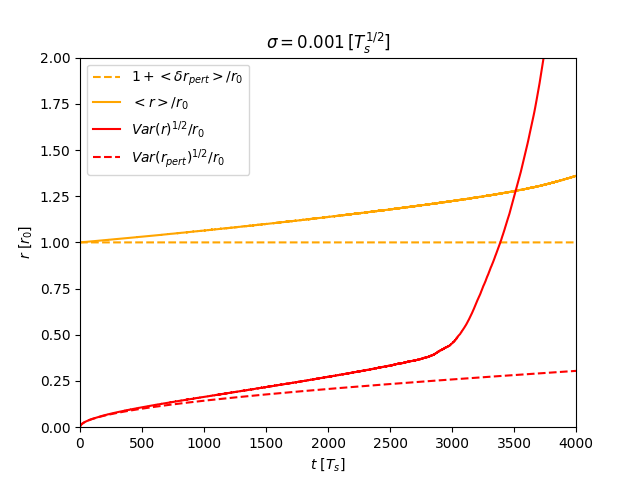}
\includegraphics[width=0.49\linewidth]{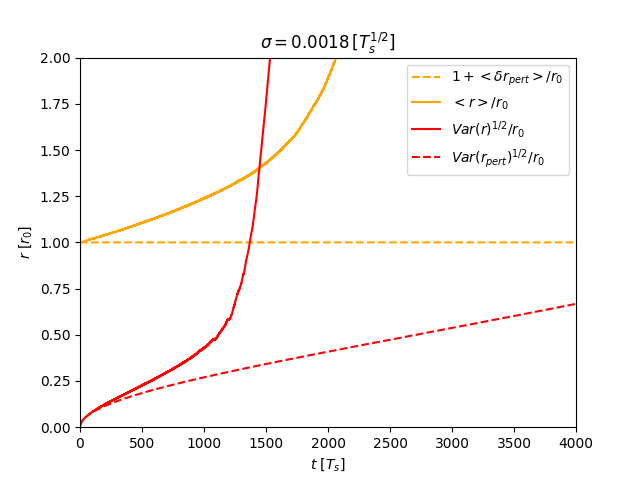}
\includegraphics[width=0.49\linewidth]{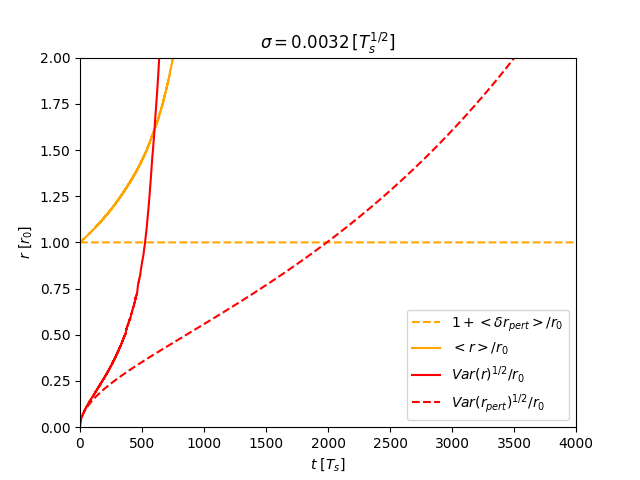}
\end{center}
\caption{Mean (orange) and variance (red) of the relative deviation of the radial distance, $\delta r/r_0$, for different values of $\sigma$ chosen nonrealistically to exaggerate the effect for visualization. The timescale is given by the orbital period $T_s = 2\pi \, \bar{\Omega}_0^{-1}$. Solid lines are computed over samples of numerical realizations with at least $N = 1.5 \times 10^4$, while dashed lines are the analytical expectations computed from the linearized equations. The initial conditions have been set to $\delta r(0) = 0$ and $\delta \dot{r}(0) = 0$ all around, which emphasizes the role of the additive noise in pushing the system out of equilibrium.
}
\label{fig:plot-binary}
\end{figure}

Note that, unlike for the case of nonstochastic time-dependent $G$ described by Eq.\eqref{time-dep-G}, in which the deviation $\delta r$ is linear in $t$  \cite{Nordtvedt1999}, here the deviation is proportional to $\sqrt t$. However, at longer times the effect is more dramatic. Although the validity of this calculation is restricted [because we neglected nonlinear terms in Eq.\eqref{orbital-pert}], the numerical simulations show that the nonlinear terms enhance the variance even further, as exhibited in Fig.~\ref{fig:plot-binary} for different values of $\sigma$. Eventual dissipative effects associated with the fundamental origin of the stochastic fluctuations may potentially curb this growth and mitigate the observable effects. One could incorporate this phenomenologically by adding a term $\beta \, \delta\dot{r}$ to, for example, Eq.~\eqref{orbital-pert-linear}. However, this would not be very illuminating without knowledge of the fundamental theory to properly connect this new phenomenological parameter $\beta$ with the noise amplitude $\sigma$ by means of a fluctuation-dissipation theorem.

While laser ranging is available for bodies orbiting the Earth, like the Moon or artificial satellites, for other binary systems  the observed orbital elements are the period, the eccentricity, the inclination, etc. A detailed analysis of the influence of stochastic gravitational waves on the orbital parameters, including resonant effects, has been presented in Ref.~\cite{Blas:2021mpc}. A similar analysis could be performed  for the case in which  the source of stochasticity is the time dependence of the gravitational coupling discussed in this paper. For instance, for the simplified approach considered here one could relate the stochastic variations of $r(t)$ with those of the angular velocity and of the orbital period through the conservation of angular momentum.

\section{Discussion} \label{sec:disc}

In this paper we have studied the effects of a purely time-dependent stochastic contribution to the gravitational coupling $G$. Following Ref.~\cite{deCesare:2016dnp} we have first considered the cosmological implications by studying the Friedmann-Langevin equations without a cosmological constant. We have extended what was presented there by computing also the stochastic mean value and variance of the Hubble parameter in a perturbative approximation valid at early times. We have compared our analytic results with numerical simulations in the case of white noise, showing good agreement within the regime of validity of our approximation. Our results show that $\mathcal{O}(1)$ deviations from the deterministic matter-dominated evolution of the Hubble parameter are to be expected after a characteristic time $T_{\text{NP}}$ given in Eq.~\eqref{t_NP-cosmo}. In order to provide a resolution to the cosmological constant problem this must happen, taking $H_D^{(i)} \sim H_0$, i.e., the Hubble constant today, on a timescale shorter than the age of the Universe. This condition imposes a lower bound for the noise amplitude $\sigma$ of 
\begin{equation} \label{cosmo-lower-bound}
    \sigma_{\text{cosmo}} \gtrsim 10^5 \, \text{yr}^{1/2} 
    \begin{cases}
        1 \quad & \alpha \lesssim 1, \\
        \alpha^{-1/2} \quad & \alpha \gg 1.
    \end{cases}
\end{equation}
Nevertheless, this does not guarantee that the Hubble parameter will stabilize at a positive constant value. The numerical simulations exhibit a strong dependence of the results with the stochastic initial conditions assumed for the Hubble parameter. For a sufficiently wide positive half-normal distribution ($\alpha \gg 1$), the solutions at large times look as though they are dominated by a positive cosmological constant, as described in Ref.~\cite{deCesare:2016dnp}.

Given the dependence with the initial conditions, the scenario where stochastic variations of $G$ produce an effective evolution of the Hubble parameter similar to that produced by a cosmological constant cannot be put to the test. This is due to the stochastic nature of the process and the fact that we can only observe one realization of it at cosmological scales. The situation could be improved by studying the effects of this scenario at shorter scales where one can rely on statistical analysis over a large number of systems. In the case of binary systems at astrophysical scales, this would benefit from ongoing precision observations of pulsars and also within the Solar System. 

We have then moved on to study the effects that this kind of stochasticity in the gravitational coupling has on binary systems. We have done so by looking at the effects of noise on the perturbations around circular orbits, once again computing their stochastic mean value and variance. We have found a combined effect of multiplicative and additive noise that is similar to parametric resonance, but that is larger and independent of initial conditions. We have then once again compared with numerical simulations in the case of white noise. Our main result for binary systems can be summarized as follows: at the linearized level, stochastic fluctuations of $G$ produce a cumulative effect on the distance between bodies. At short times, the variance  grows as $\sqrt t$. At larger times, numerical simulations suggest that the growth of the variance is exponential. 

Laser ranging can be used to put bounds on the amplitude of the frequency spectrum of the noise, Eq.~\eqref{LLR-bound-3}, of order
\begin{equation*}
    \sigma_{\text{LLR}} \lesssim 10^{-13} \, \text{yr}^{1/2},
\end{equation*}
which immediately seems in strong contradiction with the cosmologically motivated value of Eq.~\eqref{cosmo-lower-bound}, unless one allows for unnaturally large stochastic initial fluctuations in cosmology, i.e., $\alpha \gtrsim 10^{36}$ (which imply a cutoff still lower than $M_\text{P}$). As one might expect though, this approach can only constrain specific ranges of the noise power spectrum $S(\omega)$ to which the studied binary systems are sensitive. This can only be extrapolated to the infrared cosmological scales under some assumption regarding the tilt of the spectrum. Indeed, assuming a power-law scaling
\begin{equation}
    S(\omega) \sim \omega^{-p},
\end{equation}
it is possible to bridge the gap between the Solar System and cosmological scales with $p \gtrsim 3$.

Our results can be generalized in any number of ways. In the context of the mergers of black holes and other compact objects, this type of stochastic effect may impact not only the binary dynamics but also the generation of gravitational waves. This might be of interest in light of future next-generation gravitational-wave observatories. In the context of nonlocal gravity, it has been pointed out \cite{Belgacem:2018wtb} that the limit on $\dot G/G$ from LLR can be used to rule out some models not compatible with it. It would be interesting to extend that analysis to the case in which $G$ has stochastic fluctuations and dissipation, since both effects are expected when the nonlocalities are induced by the integration of quantum gravitational degrees of freedom. 

Finally, we would like to stress that the MSA approach described in this paper could also be used to analyze eventual resonant effects in the studies of the stochastic corrections induced by gravitons on geodesics deviation and tidal forces which, up to now, have been addressed perturbatively in the amplitude of the noise \cite{Parikh:2020nrd, Parikh:2020fhy,Haba:2020jqs,Cho:2021gvg,Cho:2023dmh,Chawla:2021lop}.  

\section*{Acknowledgments}
The work of L.G.T. was supported by the Czech Science Foundation (Grant No. 20-28525S) and by European Union (Grant No. 101063210). The work of F.D.M. is supported by Agencia Nacional de Promoci\'on Cient\'ifica y Tecnol\'ogica (ANPCyT), Consejo Nacional de Investigaciones Cient\'ificas y T\'ecnicas (CONICET), and Universidad Nacional de Cuyo (UNCuyo).

\section*{Appendix}
 In this appendix we provide some details of the MSA of the stochastic differential equation \eqref{MSA-eq}. As described in the text, the functions $A(\tau)$ and $B(\tau)$ are determined by the condition that $\langle Z_2(t,\tau)\rangle$ does not have secular terms. 

From Eq.\eqref{eq:Z2} we obtain
 \begin{equation} \label{Z2-mean}
\begin{split}
    \mv{Z_2(t,\tau)} = &
    \int_0^t dt' \sin[\omega(t-t')] \bigg\{ -2 i \left[ A'(\tau) e^{i\omega t'} - B'(\tau) e^{-i\omega t'} \right] \\
    & + 4\omega \int_0^{t'} dt'' \sin[\omega (t'-t'')] R(t'-t'') \left[ A(\tau) e^{i\omega t''} + B(\tau) e^{-i\omega t''} + \gamma \right] \bigg\},
\end{split}
\end{equation}
where we have used Eq.~\eqref{correlation}. Secular terms are those that can grow as $t$, and in the previous expression we can see that these can occur when there are contributions inside the integral over $t'$ that can resonate with frequency $\omega$, i.e., terms like $e^{i\omega t'}$ or $e^{-i\omega t'}$. In order to identify these, we need to massage the inner integral a little bit. After a change of variables $u = t' - t''$, we find integrals of the form
\begin{equation}
    \int_{0}^{\infty} du \, \sin(\omega u) R(u) \, e^{i\omega u} = \frac{i}{2} \left[S(0) - S(2\omega) \right],
\end{equation}
and its complex conjugate, where the integration has been extended to infinity ($t' \to \infty$) since the secular effects are associated to times much longer than $\omega^{-1}$ as discussed before, allowing us to recast it in terms of the Fourier transform $S(\omega)$ of the correlation function, as defined in Eq.~\eqref{Fourier-correlation}. 

At this stage we implement the resummation of the secular terms by imposing that the quantity in curly brackets in Eq.~\eqref{Z2-mean} has no such resonant terms, which will impose conditions on $A(\tau)$ and $B(\tau)$ in the form of a pair of ordinary differential equations in $\tau$,
\begin{equation} \label{A-B-eqs}
\begin{cases} 
    A'(\tau) + \omega \left[S(0) - S^*(2\omega) \right] A(\tau) = 0, \\
    B'(\tau) + \omega \left[S(0) - S(2\omega) \right] B(\tau) = 0 .
\end{cases}
\end{equation}
These are immediately solved as 
\begin{equation} \label{A-B-mean}
\begin{cases} 
    A(\tau) = A(0) \, e^{\left[S^*(2\omega) - S(0) \right] \omega \tau},  \\
    B(\tau) = B(0) \, e^{\left[S(2\omega) - S(0) \right] \omega \tau}   .
\end{cases}
\end{equation}
From these equations it is easy to compute $\langle z(t)\rangle$.

We now consider the evaluation of $\langle z(t)^2\rangle$.
We only need to compute $\langle Z_1(t,\tau)^2 \rangle$. Using Eq.~\eqref{Z1}, we obtain
\begin{eqnarray} \label{Z1Z1-mean}
    \langle Z_1(t,\tau)^2 \rangle &=& \left[C(\tau) e^{i\omega t} + D(\tau) e^{-i\omega t}\right]^2 + 4 \omega \int_0^t dt' \sin[\omega (t-t')] \int_0^t dt'' \sin[\omega (t-t'')] \notag\\
    &&\times R(t'-t'') \left[ A(\tau) e^{i\omega t'} + B(\tau) e^{-i\omega t'} + \gamma \right] \left[ A(\tau) e^{i\omega t''} + B(\tau) e^{-i\omega t''} + \gamma \right].    
\end{eqnarray}
In order to find the secular terms here and combine them with those coming from $2 Z_0(t,\tau) \langle Z_2(t,\tau) \rangle$, we need to recast this expression in a way that resembles Eq.~\eqref{Z2-mean}, that is, with two nested integrals. For this we use that, for any symmetric function $f(t',t'') = f(t'',t')$, 
\begin{equation}
    \int_0^t dt' \int_0^t dt'' \, f(t',t'') = 2 \int_0^t dt' \int_0^{t'} dt'' \, f(t',t'').
\end{equation}  
Then, Eq.~\eqref{Z1Z1-mean} reads
\begin{eqnarray} \label{Z1Z1-mean2}
    \langle Z_1(t,\tau)^2 \rangle &=&  8 \omega \int_0^t dt' \sin[\omega (t-t')] \left[ A(\tau) e^{i\omega t'} + B(\tau) e^{-i\omega t'} + \gamma \right] \notag\\
    &&\times \int_0^{t'} dt'' \sin[\omega (t-t'')] R(t'-t'') \left[ A(\tau) e^{i\omega t''} + B(\tau) e^{-i\omega t''} + \gamma \right].    
\end{eqnarray}
Notice that this expression is similar to Eq.~\eqref{Z2-mean} with an important difference, as the inner integral after a change of variables is now of the form
\begin{equation}
    \int_{0}^{\infty} du \, \sin[\omega (t-t') + \omega u] R(u) \, e^{i\omega u} = \frac{i}{2} \left[ e^{-i\omega(t-t')} S(0) - e^{i\omega(t-t')} S(2\omega) \right],
\end{equation}
and its complex conjugate. Once again we have extended the limit of integration $t' \to \infty$.

Finally, combining Eqs.~\eqref{Z0}, \eqref{Z2-mean}, and \eqref{Z1Z1-mean2} to form Eq.~\eqref{second-order-zz}, and then following the same procedure to get from Eq.~\eqref{Z2-mean} to Eq.~\eqref{A-B-eqs}, we now obtain instead that the secular terms vanish if
\begin{equation} \label{AA-AB-BB-eqs}
\begin{cases} 
    [A(\tau)^2]' + 2 \omega \left[2 S(0) - S^*(2\omega) \right] A(\tau)^2 = 0, \\    
    [B(\tau)^2]' + 2 \omega \left[2 S(0) - S(2\omega) \right] B(\tau)^2 = 0, \\
    [2 A(\tau) B(\tau)]' - 4 \omega \, \text{Re}[S(2\omega)] (2 A(\tau) B(\tau)) = 4 \gamma^2 \omega \, \text{Re}[S(\omega)],
\end{cases}
\end{equation}
which, as previously discussed, are to be solved independently for $A(\tau)^2$, $B(\tau)^2$, and $2 A(\tau) B(\tau)$. The solutions for $A(\tau)^2$, $B(\tau)^2$ are again very simple
\begin{equation} \label{AA-BB-mean2}
\begin{cases} 
    [A(\tau)]^2 = A(0)^2 \, e^{2 \left[S^*(2\omega) - 2S(0) \right] \omega \epsilon^2 t},  \\
    [B(\tau)]^2 = B(0)^2 \, e^{2 \left[S(2\omega) - 2S(0) \right] \omega \epsilon^2 t} ,
\end{cases}
\end{equation}
and are related to the oscillatory part of Eq.~\eqref{Z0Z0}. On the other hand, for the nonoscillatory part, the third equation in \eqref{AA-AB-BB-eqs} has a source term proportional to $\gamma^2$. This is where we see the effect of the additive noise appearing. The solution has both a homogeneous part and a particular part, and is given by
\begin{equation}\label{2AB-app}
    [2 A(\tau) B(\tau)] = 2 A(0) B(0) \, e^{4 \omega \text{Re}[S(2\omega)] \epsilon^2 t} + \gamma^2 \, \frac{\text{Re}[S(\omega)]}{\text{Re}[S(2\omega)]} \left( e^{4 \omega \text{Re}[S(2\omega)] \epsilon^2 t} - 1 \right).
\end{equation}
Equations \eqref{AA-BB-mean2} and \eqref{2AB-app} are summarized in Eq.\eqref{AA-BB-mean} of Sec.~\ref{sec:MSA}.

\bibliographystyle{unsrt}
\bibliography{Stochastic-G}

\begin{thebibliography}{10}

\bibitem{Carney:2018ofe}
Daniel Carney, Philip C.~E. Stamp, and Jacob~M. Taylor.
\newblock {Tabletop experiments for quantum gravity: a user\textquoteright{}s
  manual}.
\newblock {\em Class. Quant. Grav.}, 36(3):034001, 2019.

\bibitem{Donoghue:1994dn}
John~F. Donoghue.
\newblock {General relativity as an effective field theory: The leading quantum
  corrections}.
\newblock {\em Phys. Rev. D}, 50:3874--3888, 1994.

\bibitem{Donoghue:2022eay}
John~F. Donoghue.
\newblock {Quantum General Relativity and Effective Field Theory}.
\newblock arXiv 2211.09902, 2022.

\bibitem{Birrell:1982ix}
N.~D. Birrell and P.~C.~W. Davies.
\newblock {\em {Quantum Fields in Curved Space}}.
\newblock Cambridge Monographs on Mathematical Physics. Cambridge Univ. Press,
  Cambridge, UK, 2 1984.

\bibitem{Parker:2009uva}
Leonard~E. Parker and D.~Toms.
\newblock {\em {Quantum Field Theory in Curved Spacetime}: {Quantized Fields
  and Gravity}}.
\newblock Cambridge Monographs on Mathematical Physics. Cambridge University
  Press, 8 2009.

\bibitem{Hu:1991di}
B.~L. Hu, Juan~Pablo Paz, and Yu-hong Zhang.
\newblock {Quantum Brownian motion in a general environment: 1. Exact master
  equation with nonlocal dissipation and colored noise}.
\newblock {\em Phys. Rev. D}, 45:2843--2861, 1992.

\bibitem{Hu:2020luk}
Bei-Lok~B. Hu and Enric Verdaguer.
\newblock {\em {Semiclassical and Stochastic Gravity}: {Quantum Field Effects
  on Curved Spacetime}}.
\newblock Cambridge Monographs on Mathematical Physics. Cambridge University
  Press, Cambridge, 1 2020.

\bibitem{Parikh:2020nrd}
Maulik Parikh, Frank Wilczek, and George Zahariade.
\newblock {The Noise of Gravitons}.
\newblock {\em Int. J. Mod. Phys. D}, 29(14):2042001, 2020.

\bibitem{Parikh:2020fhy}
Maulik Parikh, Frank Wilczek, and George Zahariade.
\newblock {Signatures of the quantization of gravity at gravitational wave
  detectors}.
\newblock {\em Phys. Rev. D}, 104(4):046021, 2021.

\bibitem{Haba:2020jqs}
Z.~Haba.
\newblock {State-dependent graviton noise in the equation of geodesic
  deviation}.
\newblock {\em Eur. Phys. J. C}, 81(1):40, 2021.

\bibitem{Cho:2021gvg}
Hing-Tong Cho and Bei-Lok Hu.
\newblock {Quantum noise of gravitons and stochastic force on geodesic
  separation}.
\newblock {\em Phys. Rev. D}, 105(8):086004, 2022.

\bibitem{Cho:2023dmh}
Hing-Tong Cho and Bei-Lok Hu.
\newblock {Graviton noise on tidal forces and geodesic congruences}.
\newblock arXiv 2301.06325, 2023.

\bibitem{Chawla:2021lop}
Samarth Chawla and Maulik Parikh.
\newblock {Quantum Gravity Corrections to the Fall of the Apple}.
\newblock arXiv 2112.14730, 2021.

\bibitem{Dalvit:1997yc}
Diego A.~R. Dalvit and Francisco~D. Mazzitelli.
\newblock {Geodesics, gravitons and the gauge fixing problem}.
\newblock {\em Phys. Rev. D}, 56:7779--7787, 1997.

\bibitem{Dalvit:1999wd}
Diego A.~R. Dalvit and Francisco~D. Mazzitelli.
\newblock {Quantum corrected geodesics}.
\newblock {\em Phys. Rev. D}, 60:084018, 1999.

\bibitem{dePaulaNetto:2021axj}
Tib\'erio de~Paula~Netto, Leonardo Modesto, and Ilya~L. Shapiro.
\newblock {Universal leading quantum correction to the Newton potential}.
\newblock {\em Eur. Phys. J. C}, 82(2):160, 2022.

\bibitem{Wang:2017oiy}
Qingdi Wang, Zhen Zhu, and William~G. Unruh.
\newblock {How the huge energy of quantum vacuum gravitates to drive the slow
  accelerating expansion of the Universe}.
\newblock {\em Phys. Rev. D}, 95(10):103504, 2017.

\bibitem{Lozano:2020xga}
Ezequiel Lozano and Francisco~Diego Mazzitelli.
\newblock {The role of noise in the early universe}.
\newblock {\em Int. J. Mod. Phys. D}, 30(15):2150117, 2021.

\bibitem{Calzetta:2008iqa}
Esteban~A. Calzetta and Bei-Lok~B. Hu.
\newblock {\em {Nonequilibrium Quantum Field Theory}}.
\newblock Oxford University Press, 2009.

\bibitem{Belgacem:2017cqo}
Enis Belgacem, Yves Dirian, Stefano Foffa, and Michele Maggiore.
\newblock {Nonlocal gravity. Conceptual aspects and cosmological predictions}.
\newblock {\em JCAP}, 03:002, 2018.

\bibitem{deCesare:2016dnp}
Marco de~Cesare, Fedele Lizzi, and Mairi Sakellariadou.
\newblock {Effective cosmological constant induced by stochastic fluctuations
  of Newton's constant}.
\newblock {\em Phys. Lett. B}, 760:498--501, 2016.

\bibitem{Murphy_2013}
T~W Murphy.
\newblock Lunar laser ranging: the millimeter challenge.
\newblock {\em Reports on Progress in Physics}, 76(7):076901, 2013.

\bibitem{Will:2014kxa}
Clifford~M. Will.
\newblock {The Confrontation between General Relativity and Experiment}.
\newblock {\em Living Rev. Rel.}, 17:4, 2014.

\bibitem{Blas:2021mpc}
Diego Blas and Alexander~C. Jenkins.
\newblock {Detecting stochastic gravitational waves with binary resonance}.
\newblock {\em Phys. Rev. D}, 105(6):064021, 2022.

\bibitem{Deprit}
Andr{\'e} Deprit.
\newblock The secular acceleratons in gylden's problem.
\newblock {\em Celestial mechanics}, 31(1):1--22, 1983.

\bibitem{Abad_2020}
Alberto Abad, Manuel Calvo, Jos{\'e}~A. Docobo, and Antonio Elipe.
\newblock On the orbital elements of the two-body problem with slowly
  decreasing mass: The gyld{\'e}n--mestchersky cases.
\newblock {\em The Astronomical Journal}, 160(5):203, oct 2020.

\bibitem{bender78:AMM}
C.~M. Bender and S.~A. Orszag.
\newblock {\em {Advanced Mathematical Methods for Scientists and Engineers}}.
\newblock McGraw-Hill, 1978.

\bibitem{Fritzsch:2012qc}
Harald Fritzsch and Joan Sola.
\newblock {Matter Non-conservation in the Universe and Dynamical Dark Energy}.
\newblock {\em Class. Quant. Grav.}, 29:215002, 2012.

\bibitem{Muller:2007zzb}
Jurgen Muller and Liliane Biskupek.
\newblock {Variations of the gravitational constant from lunar laser ranging
  data}.
\newblock {\em Class. Quant. Grav.}, 24:4533--4538, 2007.

\bibitem{Merkowitz:2010kka}
Stephen~M. Merkowitz.
\newblock {Tests of Gravity Using Lunar Laser Ranging}.
\newblock {\em Living Rev. Rel.}, 13:7, 2010.

\bibitem{Biskupek:2020fem}
Liliane Biskupek, J\"urgen M\"uller, and Jean-Marie Torre.
\newblock {Benefit of New High-Precision LLR Data for the Determination of
  Relativistic Parameters}.
\newblock {\em Universe}, 7(2):34, 2021.

\bibitem{Nordtvedt1999}
Kenneth Nordtvedt.
\newblock 30 years of lunar laser ranging and the gravitational interaction.
\newblock {\em Classical and Quantum Gravity}, 16(12A):A101, dec 1999.

\bibitem{papanicolau1971stochastic}
George Papanicolau and Joseph~B Keller.
\newblock Stochastic differential equations with applications to random
  harmonic oscillators and wave propagation in random media.
\newblock {\em SIAM Journal on Applied Mathematics}, 21(2):287--305, 1971.

\bibitem{PhysRevE.48.4309}
Jaume Masoliver and Josep~M. Porr\`a.
\newblock Harmonic oscillators driven by colored noise: Crossovers, resonances,
  and spectra.
\newblock {\em Phys. Rev. E}, 48:4309--4319, Dec 1993.

\bibitem{Mantinan:2022vcn}
Mat\'\i{}as Manti\~nan, Francisco~D. Mazzitelli, and Leonardo~G. Trombetta.
\newblock {Stochastic Particle Creation: From the Dynamical Casimir Effect to
  Cosmology}.
\newblock {\em Entropy}, 25(1):151, 2023.

\bibitem{Murphy_2012}
T~W Murphy, E~G Adelberger, J~B~R Battat, C~D Hoyle, N~H Johnson, R~J McMillan,
  C~W Stubbs, and H~E Swanson.
\newblock Apollo: millimeter lunar laser ranging.
\newblock {\em Classical and Quantum Gravity}, 29(18):184005, aug 2012.

\bibitem{Belgacem:2018wtb}
Enis Belgacem, Andreas Finke, Antonia Frassino, and Michele Maggiore.
\newblock {Testing nonlocal gravity with Lunar Laser Ranging}.
\newblock {\em JCAP}, 02:035, 2019.

\end{thebibliography}

\end{document}